\UseRawInputEncoding
\documentclass[a4paper,fleqn,usenatbib]{mnras}

\usepackage{subfigure}
\usepackage{epsfig}
\usepackage{rotating}
\usepackage[T1]{fontenc}
\usepackage{ae,aecompl}
\usepackage{tikz}

\usepackage{graphicx}	
\usepackage{amsmath}	
\usepackage{amssymb}	

\usepackage{color}
\usepackage{natbib}

\usepackage{booktabs}

\newcommand{\Ms}{{\ensuremath{\mathrm{M}_{\odot} }}}

\title[Pop III stars]{Massive black holes or stars first: the key is the residual cosmic electron fraction}

\author[Latif and Whalen]{Muhammad A. Latif$^{1}$\thanks{E-mail: latifne@gmail.com}, 
Sadegh Khochfar$^{2}$
\\
\\
$^{1}$ Physics Department, College of Science, United Arab Emirates University, PO Box 15551, Al-Ain, UAE \\ 
$^{2}$Institute for Astronomy, University of Edinburgh, Royal Observatory, Blackford Hill, Edinburgh EH9 3HJ, UK  \\
}

\date{Accepted XXX. Received YYY; in original form ZZZ}

\pubyear{2024}

\begin{document}
\label{firstpage}
\pagerange{\pageref{firstpage}--\pageref{lastpage}}
\maketitle

\begin{abstract}

Recent James Webb Space Telescope observations have unveiled that the first supermassive black holes (SMBHs) were in place at z $\geq$ 10, a few hundred Myrs after the Big Bang. These discoveries are providing strong constraints on the seeding of BHs and the nature of the first objects in the Universe.  
Here, we study the impact of the freeze-out electron fractions ($f_e$) at the end of the epoch of cosmic recombination on the formation of the first structures in the Universe. 
At $f_e$ below the current fiducial cosmic values of $\rm \sim 10^{-4}$,  the baryonic collapse is delayed due to the lack of molecular hydrogen cooling until the host halo masses are increased by one to two orders of magnitude compared to the standard case and reach the atomic cooling limit. 
This results in an enhanced enclosed gas mass by more than an order of magnitude and higher inflow rates of up to $0.1~\Ms/{\rm yr}$.  
Such conditions are conducive to the formation of massive seed BHs with $\sim 10^{4}$ M$_{\odot}$. Our results reveal a new pathway for the formation of massive BH seeds which may naturally arise from freeze-out conditions in the early Universe. 
\end{abstract}


\section{Introduction}

 The residual electron  fraction from the epoch of recombination plays an important role in the formation of the first objects in the Universe, acting as a catalyst for the formation of molecular Hydrogen in low density gas via the reactions:
 \begin{eqnarray}
         H + e^- &\rightarrow& H^- + \gamma \\ \nonumber
    H^- + H &\rightarrow& H_2 + e^-
 \end{eqnarray}
and
 \begin{eqnarray}
         H + H^+ &\rightarrow& H_2^+ + \gamma \\ \nonumber
    H_2^+ + H &\rightarrow& H_2 + H^+
 \end{eqnarray}

In the absence of metals, $\rm H_2$ roto-vibrational modes  efficiently cool the gas down to temperatures of $\sim 200$ K and trigger collapse to densities of  $10^4$ cm$^{-3}$, at which point fragmentation starts to stall \citep{Bromm02,Yoshida08}. Numerical studies show that massive fragments of a few hundred solar masses form that can result in massive Pop III stars \citep{Latif13ApJ,Hirano14,Hosokawa16}. This simple picture however, gets complicated by the presence of magnetic fields, angular momentum and radiative feedback which lead to  potentially smaller stellar masses \citep{Clark11,Latif13ApJ,Stacy16,Sugi20,L22,R22,Latif23}. In the case that primordial gas is void or has a very small fraction of molecules, e.g. due to the Lyman-Werner radiation \citep{Latif14UV,AG15,2016MNRAS.459.4209A,Latif15a} only atomic line cooling will be effective and the associated isothermal collapse of the gas cloud at $T \sim 8000$ K will lead to the formation of a direct collapse BH  \citep{Latif13c,W19,Latif20,L22N,Latif23,L23,R24}.  

The current cosmological standard paradigm, the $\Lambda$CDM model, predicts residual $f_e$ of $\rm \sim 10^{-4}$  \citep{GP98} after the epoch of recombination. To first order the freeze out of the electrons occurs when the reaction rate for recombinations equals the expansion rate of the Universe $\Gamma_{\rm{rec}} \simeq  H(z) $ \citep{Klob90}. Expansion rates at high redshift can differ from those of the canonical $\Lambda$CDM model due to the nature of dark energy or in modified gravity theories, \citep[see for a review]{Am18}. In particular, quintessence models \citep{Cald98} such as the class of tracker models \citep{Mal02} provide a possible route to expansion rates that deviate from $\Lambda$CDM at high redshifts and track it at low redshifts. The decay of exotic primordial particles (DM or other particles) which are sometimes invoked as potential ionization sources can enhance the abundance of free electrons \citep{Be09,IO10,Old16}. Numerical studies investigating the effect of increased electron fractions in relic HII regions and under cosmic X-ray background found that higher electron fractions catalyze $\rm H_2$ formation, leading to faster cooling and an earlier baryonic collapse into dark matter halos \citep{Mac03,OS05,Hum15}.

We here present a first systematic study of the impact of lowering the residual $f_e$ on the onset of star and massive BH seed formation in the Universe. We aim to provide constraints on the critical residual $f_e$ at which the first objects preferentially are BHs instead of stars.  Our letter is structured in the following way. In section 2, we lay out the numerical methodology and simulation setup. Our main findings are presented in section 3 and we discuss our conclusions in section 4.

\section{Numerical Method}

\label{sec:method}

Our simulations are performed with the cosmological hydrodynamics code Enzo \citep{Enzo14} which uses the adaptive mesh refinement approach to refine regions of interest. They are started with cosmological initial conditions generated from the MUSIC package \citep{Hahn11} at $z$ =150 with cosmological parameters taken from the Planck 2016 data: $\Omega_{\mathrm{M}}=$ 0.308, $\Omega_{\Lambda}=$ 0.691, $\Omega_{\mathrm{b}} = $ 0.0223 and $h =$ 0.677 \citep{Planck16}. Our simulation box has a size of 0.3 cMpc/h on a side and a top grid resolution of $256^3$. We first run simulations with a uniform grid resolution of $\rm 256^3$ down to redshift 20 and select the most massive halo forming in the given volume. The selected halo is placed at the center of the box and two additional nested grids of similar resolution ($\rm 256^3$) covering 20\% of volume are employed which yield a maximum dark matter (DM) particle resolution of $\sim$ 3 $\Ms$. We further employ 18 levels of refinement during the course of the simulations which yield a maximum spatial resolution of about 40 AU. Our refinement criteria is based on the DM particle resolution, baryonic overdensity and the Jeans length, similar to \cite{L22} and \cite{L23}. The Jeans length in our simulations is resolved by at-least 32 cells. We use a non-equilibrium chemical solver \citep{Abel97} to follow the evolution of the following primordial species: $ \rm H, ~H^+,~ H^-, ~He,~ He^+, ~He^{++},~ H_2, ~H_2^+, ~e^-$ in our simulations. Our chemical model is the same as used in \cite{L22}, includes various cooling and heating processes relevant for primordial gas such as $\rm H_2$ cooling, collisional excitation and ionization cooling from H and He, recombination cooling, bremsstrahlung cooling and collisionally-induced cooling as well as heating from three-body reactions, see \cite{L22} for further details. 
In this study we consider case A recombination \citep[for a discussion of this choice see][]{Neb23}.
The chemical solver is coupled to hydrodynamics and self-consistently follows the evolution of primordial species. We ignore deuterium and its related species as they are  only relevant in HII regions or halo mergers which boost D$^+$ abundances \citep{McG08,Greif08,Bov14,Bovino14,Latif14UV}. 

We select four distinct halos taken from different  Gaussian random fields, their masses range from $\rm 1-2 \times 10^5 ~\Ms$ and collapse redshifts from 22-28. 
Each of the halos is run with different residual $f_e$ ranging from  $10^{-4}-10^{-10}$ with a decrement by a factor of ten. Our chosen range ensures to encapsulate possible values predicted by dynamical dark energy models such as the ones parametrized by a dark energy  equation of state given as \citep{2001IJMPD..10..213C,2003PhRvL..90i1301L}: 
\begin{equation}
w(z)=w_0+w_a \frac{z}{1+z}.
\end{equation}
Here, $w_0 \equiv -1$ is the value expected for a cosmological constant and $w_a$ is a scaling parameter.


\section{Results}

\begin{figure*}
\begin{center}
\includegraphics[scale=0.9]{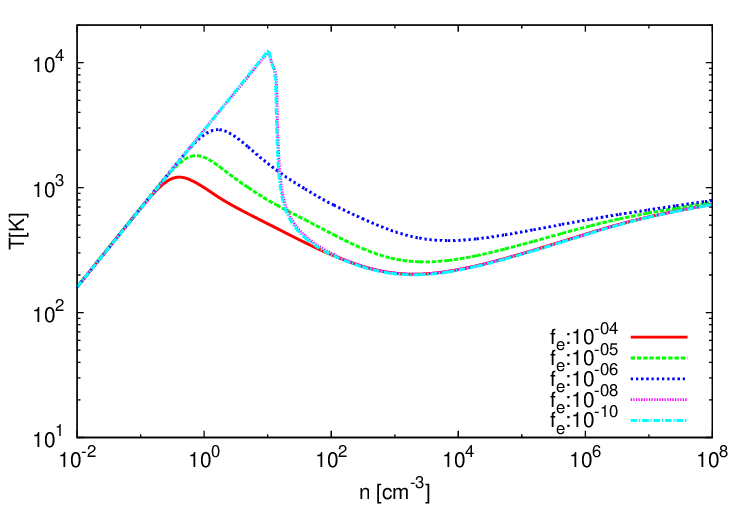}
\end{center}
\caption{Temperature vs density plot showing results for a collapsing gas cloud using the one-zone model for different initial electron fractions, see the legend.} 
\label{fig0}
\end{figure*}

To test the impact of different electron freeze-out conditions on the formation of the first objects, we first ran a number of one-zone tests \citep{Omukai01} by varying $f_e$ from $10^{-4}-10^{-10}$ and found that lowering $f_e$ resulted in delayed $\rm H_2$ formation and higher gas temperatures, see Fig. \ref{fig0}. For initial $f_e$ up to $10^{-6}$ the gas temperature rises up to 3000 K, a factor of 3 higher compared to the standard case ($f_e= 2 \times 10^{-4}$) and the cooling instability sets in at densities a factor of 3-4 higher. For $f_e = 10^{-5}$ and $f_e = 10^{-6}$, gas temperature remains about 2-4 times higher compared to the standard case. The gas temperature increases up to 13000 K, an order of magnitude higher compared to the standard case, and we also find a factor of 10 increase in the gas density for the $f_e \leq 10^{-7}$ cases. 

\begin{figure*} 
\begin{center}
\includegraphics[scale=0.9]{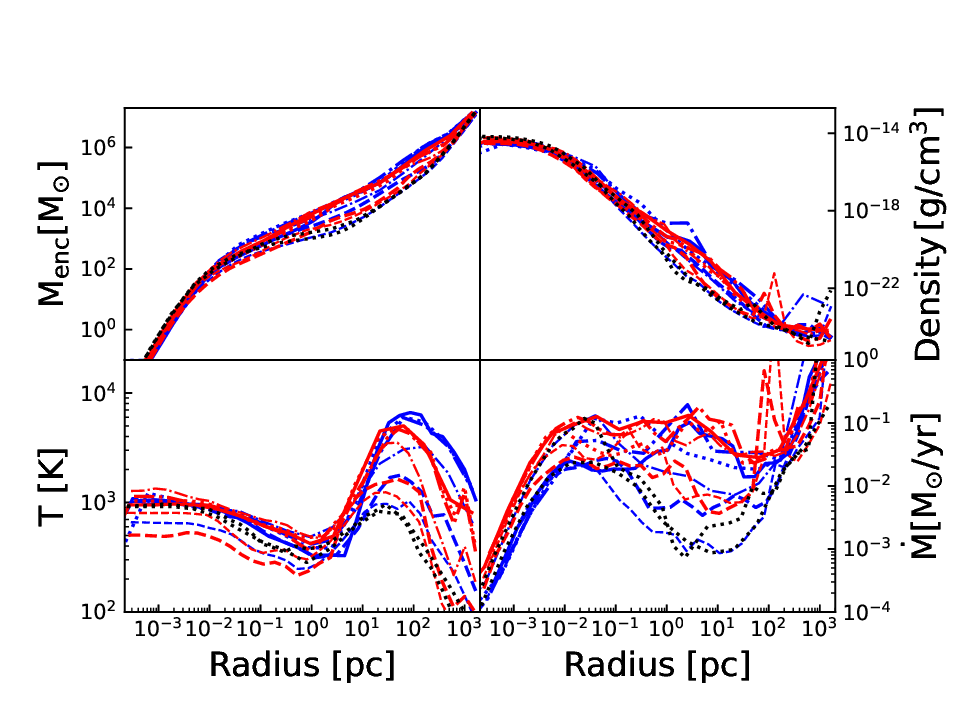}
\end{center}
\vspace{-0.1cm}
\caption{Radial profiles of gas density, temperature, enclosed baryonic mass and mass inflow rate at the end stage of simulations are shown for two representative halos. Blue lines for halo 1 and red lines for halos 2. The black dotted lines for fiducial cases ( $\rm f_e= 10^{-4}$), thin-dashed, thin dot-dashed, dot-dashed, dotted and solid lines are for $f_e$ of $\rm 10^{-5}$, $\rm 10^{-6}$,$\rm 10^{-7}$,$\rm 10^{-8}$,$\rm 10^{-9}$,$\rm 10^{-10}$, respectively. }
\label{fig1}
\end{figure*}

In total, we have performed 28 cosmological simulations for four distinct halos forming at z$>$ 20.  Their collapse redshifts (defined as the onset of baryonic collapse) for the fiducial values of $f_e$ are 22, 29, 25 and 28 for halos $1 - 4$, respectively and typical halo masses are $1-2 \times 10^5 ~\Ms$.
For the fiducial case of $f_e$, the gravitational instability is primarily driven by $\rm H_2$ cooling that kicks in at virial temperatures of about 2000 K and induces gravitational collapse, see Fig. \ref{fig1}. Same is true for the $f_e=10^{-5}$ case. However, for runs with $f_e \leq 10^{-7}$  the gravitational collapse gets delayed due to the lack of sufficient $\rm H_2$  and halo virial temperatures rise up to $\sim 10^4$ K by accretion shocks and mergers. 
Under these conditions  atomic line cooling first brings the gas temperature down to 8000 K and then baryonic collapse proceeds. At these temperatures and densities of about $\rm 10^{-24}~g/cm^3$,  $\rm H_2$ formation via the $H^-$ channel is triggered and cools the gas down to a few hundred K at smaller radii.
The same trend is observed in all halos. The central gas densities in the halos reach up to $\rm 10^{-14}~ g/cm^3$ and density profiles follow a $\rm r^{-2.2}$ power law. The bumps on the density profiles indicate the presence of gas clumps/substructures. The enclosed gas mass profile sharply increases in the central clump and then moderately increase with radius as observed in protostellar disks. Overall, the enclosed gas mass in the halos is higher for lower $f_e$ runs.

\begin{figure*} 
\begin{center}
\includegraphics[scale=0.5]{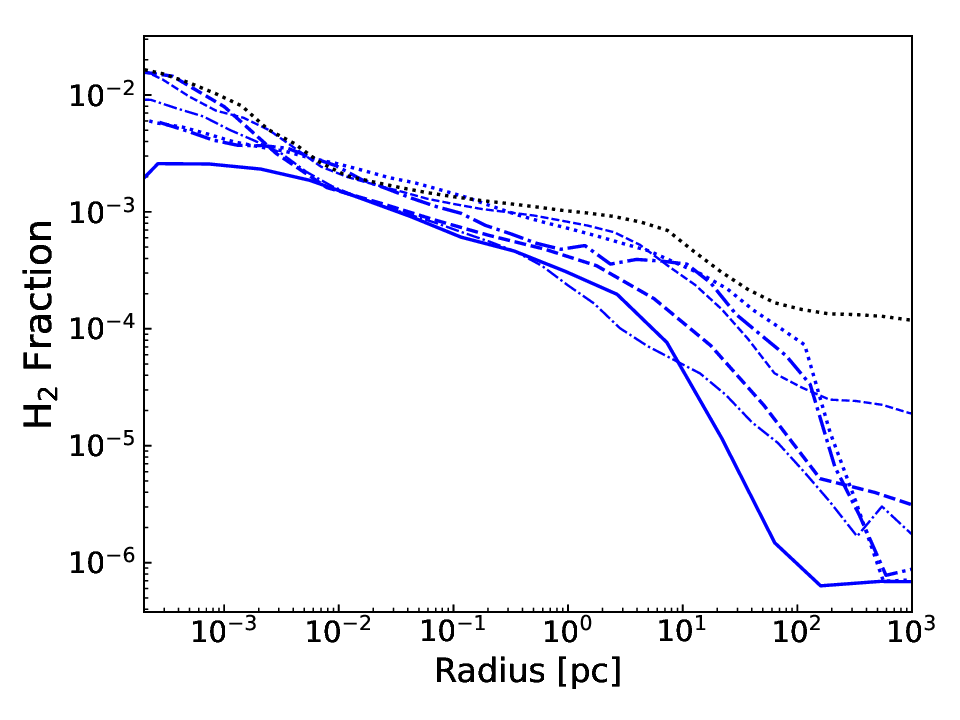}
\includegraphics[scale=0.5]{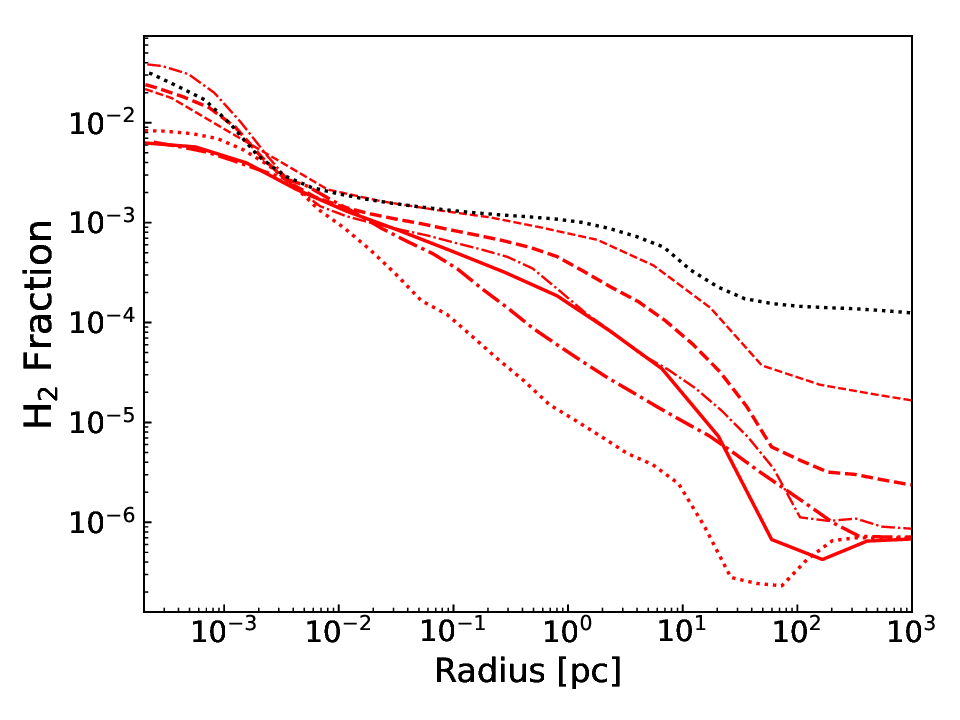}
\includegraphics[scale=0.5]{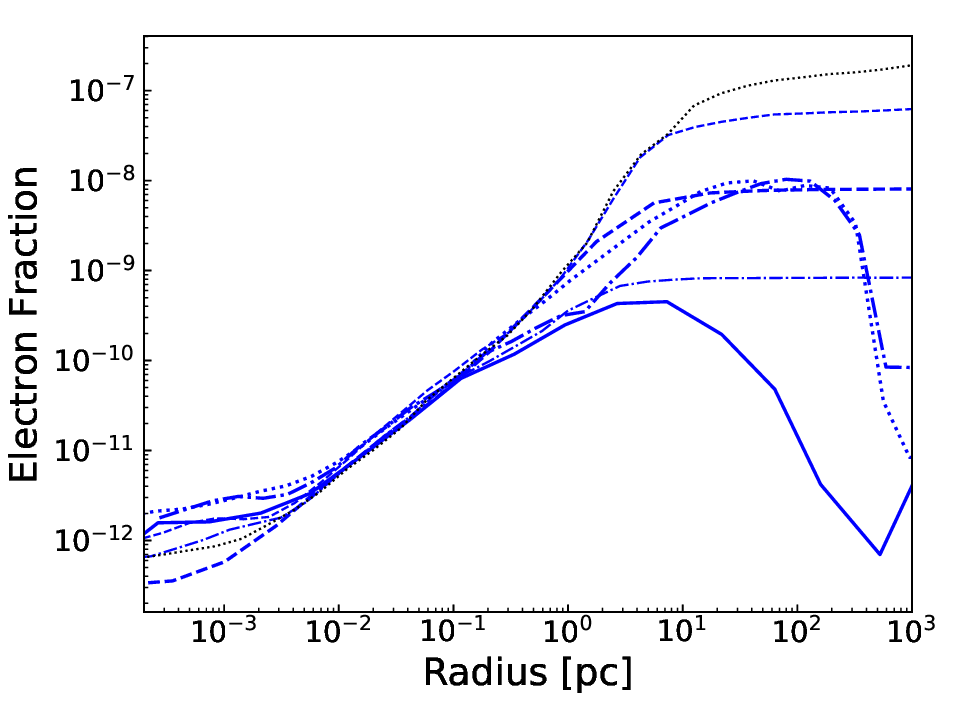}
\includegraphics[scale=0.5]{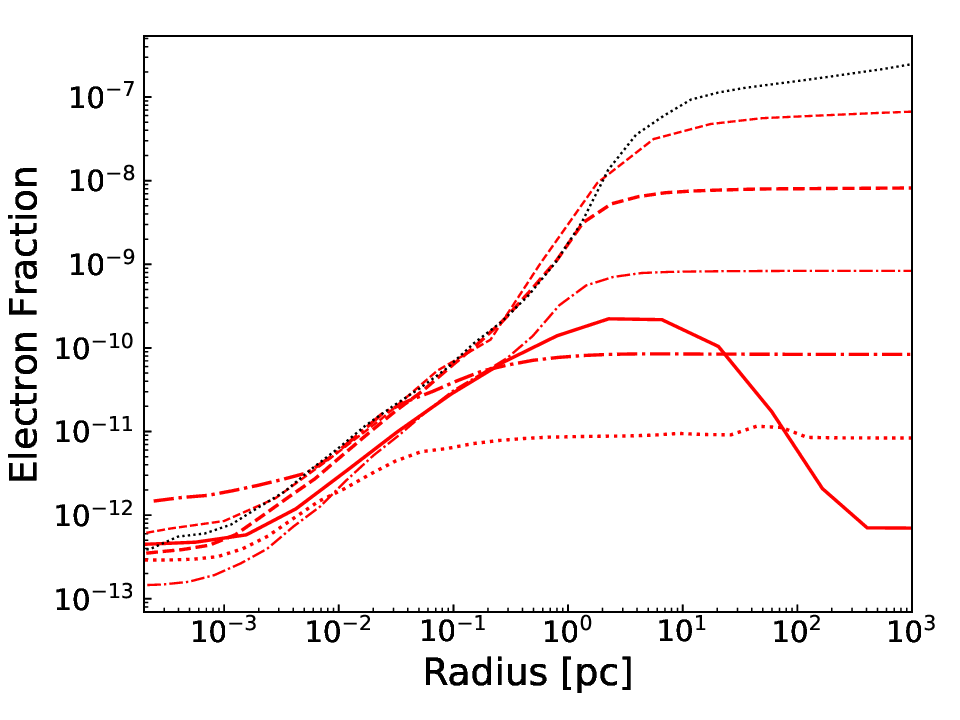}

\end{center}
\vspace{-0.1cm}
\caption{The H$_2$ and electron mass fractions at the end stage of simulations are shown for two representative halos. Blue lines for halo 1 and red lines for halos 2. The black dotted lines for fiducial cases of $\rm f_e= 10^{-4}$, thin dashed, thin dot-dashed, dot-dashed, dotted and solid lines are for $f_e$ of $\rm 10^{-5}$, $\rm 10^{-6}$, $\rm 10^{-7}$, $\rm 10^{-8}$, $\rm 10^{-9}$, $\rm 10^{-10}$, respectively.} 
\label{fig2}
\end{figure*}

The $\rm H_2$ mass fractions get boosted rapidly at radii between ten to a few hundred pc for lower $f_e$ cases, see Fig. \ref{fig2}. 
This increase in the $\rm H_2$ fractions is correlated with a higher virial temperature which catalyze $\rm H_2$ formation and boosts its abundance.
At radii between 0.001-10 pc the $\rm H_2$ abundance increases moderately and below these radii it again increases rapidly due to  three-body reactions.
In the central cloud $\rm H_2$ abundances vary from a few times $10^{-3}$ to a few times $10^{-2}$.
Overall, lower $f_e$ runs have lower $\rm H_2$ fractions at almost all radii and in spite of large differences in the initial abundances $\rm H_2$ mass fractions in the central cores differ by an order of magnitude. 
Similar trends are observed in electron abundances, they decline to smaller radii due to the recombination reactions favored for lower gas temperatures. For the lowest  $f_e=10^{-10}$ cases, electron abundances rapidly increase between 10-100 pc due to the higher temperatures which lead to collisional ionization of gas and boost electron fractions. The typical electron mass fractions in the central cloud are a few times $10^{-13}-10^{-12}$ and are inversely correlated with $\rm H_2$ abundances. Fluctuations in $f_e$ and $\rm H_2$ abundances are due to the shocks and hydrodynamical effects occurring at those radii.

\begin{figure*} 
\begin{center}
\includegraphics[scale=0.5]{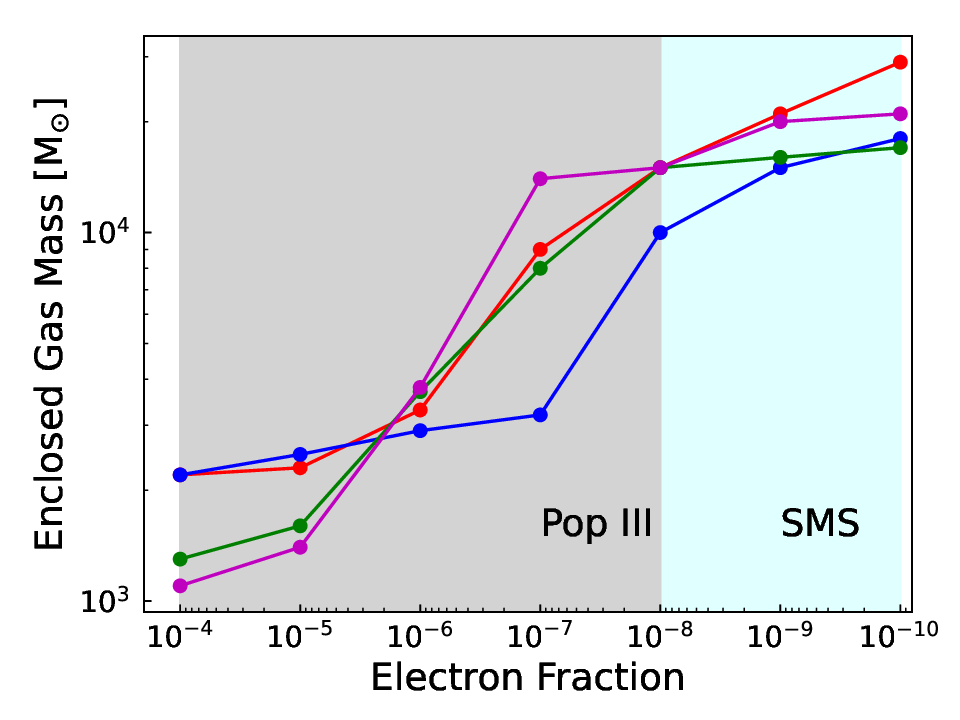}
\includegraphics[scale=0.5]{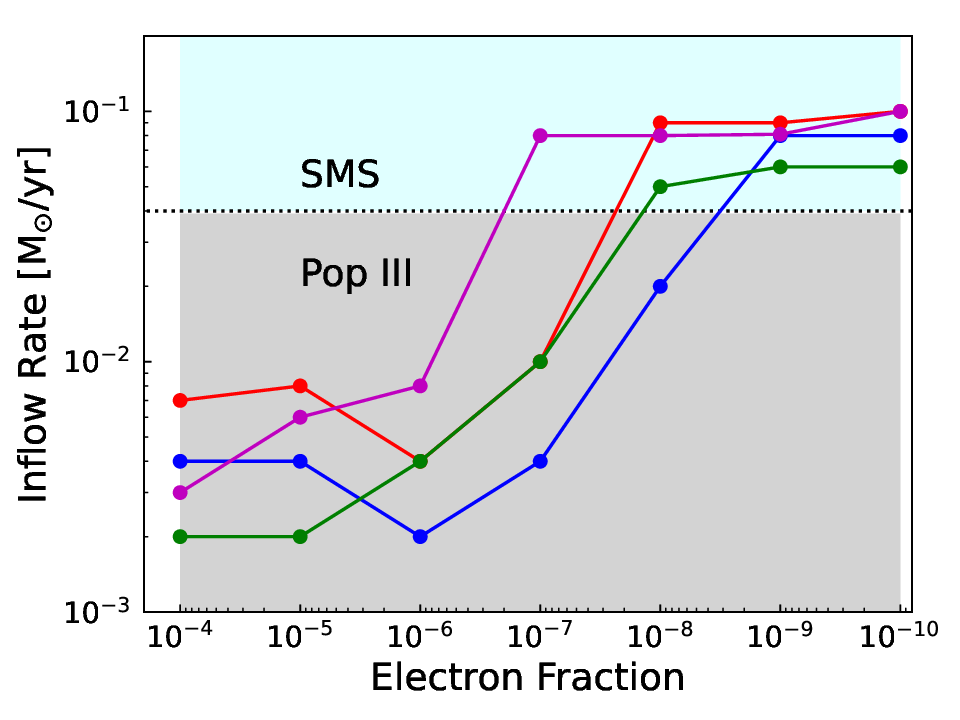}
\includegraphics[scale=0.5]{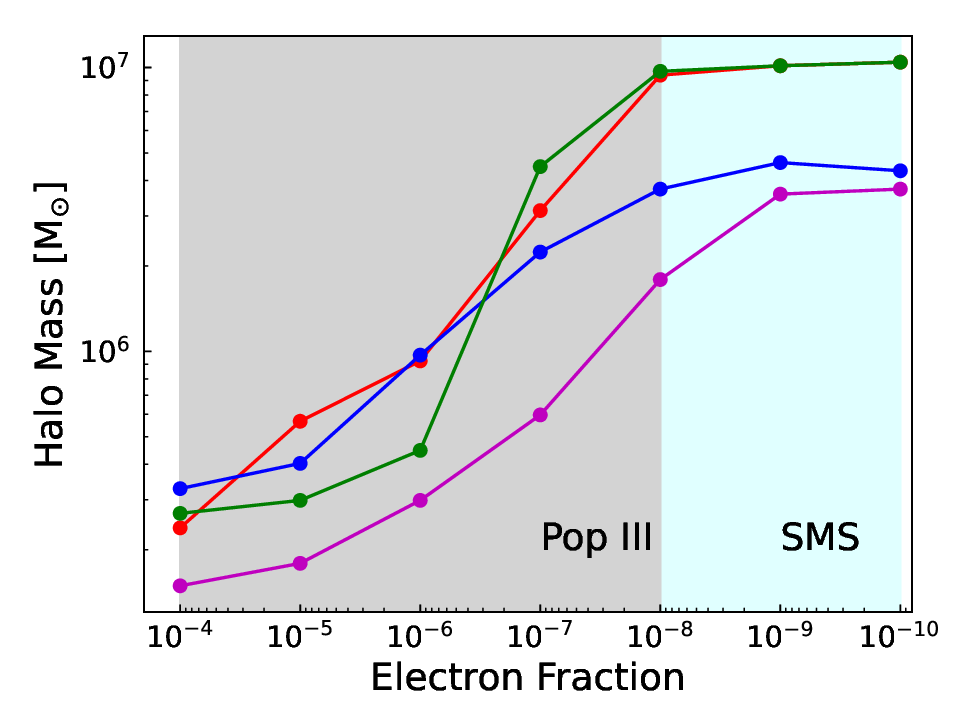} 
\includegraphics[scale=0.5]{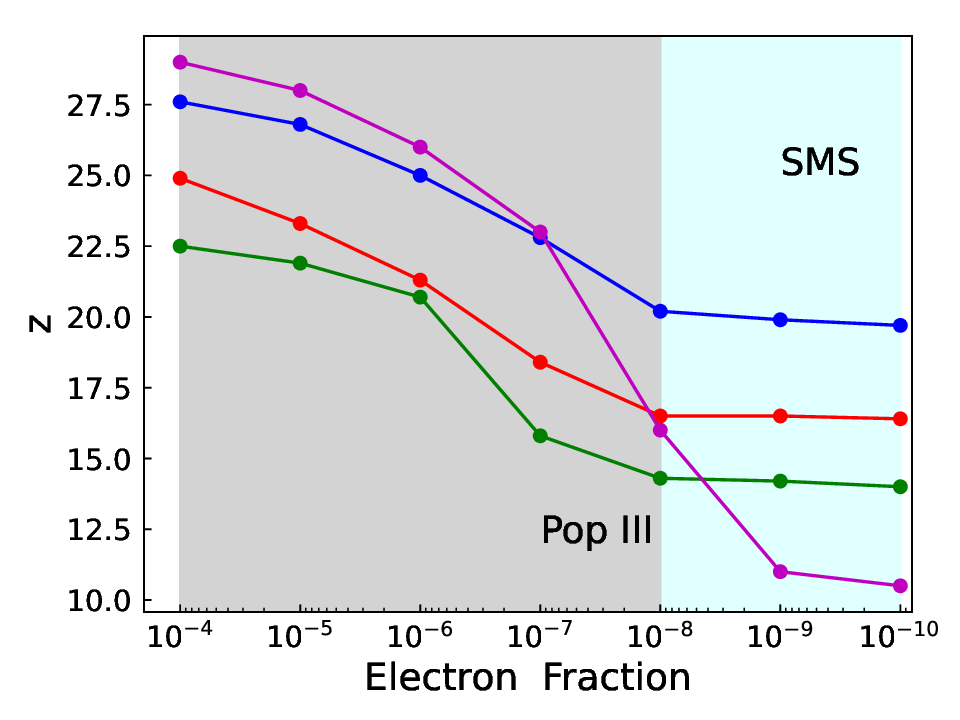}
\end{center}
\vspace{-0.1cm}
\caption{The enclosed gas mass, the mass inflow rates onto the central gas cloud, halos collapse redshifts and halo masses are shown against the initial $f_e$. The values of mass accretion rates and enclosed mass are taken at radii corresponding to the dynamical time of 2 Myr.}
\label{fig3}
\end{figure*}

To assess the amount of cold gas available for star formation in these conditions, we show the enclosed gas mass against the initial $f_e$ in Fig. \ref{fig3}. We find that typical gas masses are $\rm \sim 10^3~\Ms$ for the fiducial $f_e$ case and increase up to $\rm 3 \times 10^4~\Ms$ for the lowest $f_e=10^{-10}$ runs.
Some variations are observed from halo to halo but overall the enclosed mass increases with lowering the initial $f_e$ abundances. One of the key parameters for the fate of expected in-situ star formation is the mass inflow rate which we show in Fig. \ref{fig3}. The gas inflow rates are a few times $10^{-3}~\Ms/yr$ for the standard cases as expected in minihalos but they increase with lower initial $f_e$ and reach up to $\rm 0.1~\Ms/yr$ for $f_e=10^{-10}$ cases.
Such large inflow rates are a consequence of higher temperatures and densities in the halos and larger radial infall velocities due to the deeper potential wells.
Under these large inflow rates we expect super-massive  stars (SMSs) to form and to directly collapse into massive BHs as we discuss below \citep{Hos13,Schleicher13}. 

We show halo masses at the onset of baryonic collapse against their initial $f_e$ for all simulated cases in Fig. \ref{fig3}. For the standard $f_e$ cases, halo masses are about a few $\rm 10^5~\Ms$, they increase for lower $f_e$ runs and reach up to  $\rm 10^7~\Ms$. The halo mass becomes almost constant for $f_e \leq 10^{-8}$ and differences are mainly due to the assembly history. As expected, for low electron fractions  $f_e \leq 10^{-8}$ halo masses reach the atomic cooling limit \citep[e.g.]{2001ApJ...548..509M,Latif19}. 
Such massive and pristine halos are considered birthplaces of SMSs \citep{um16,hle17,Woods19}. 
The halo collapse redshifts vary from 22.5-28 for the standard cases. Collapse gets delayed up to a few hundred Myrs for lower $f_e$ cases and the differences in the collapse redshifts between standard and $\rm f_e=10^{-10}$ cases are $\Delta z =8.5$, $\Delta z =18$, $\Delta z =8.5$ and $\Delta z =7.9$ for halos 1-4, respectively.

\begin{figure*}
\begin{center}
\begin{tabular}{c c}
\epsfig{file=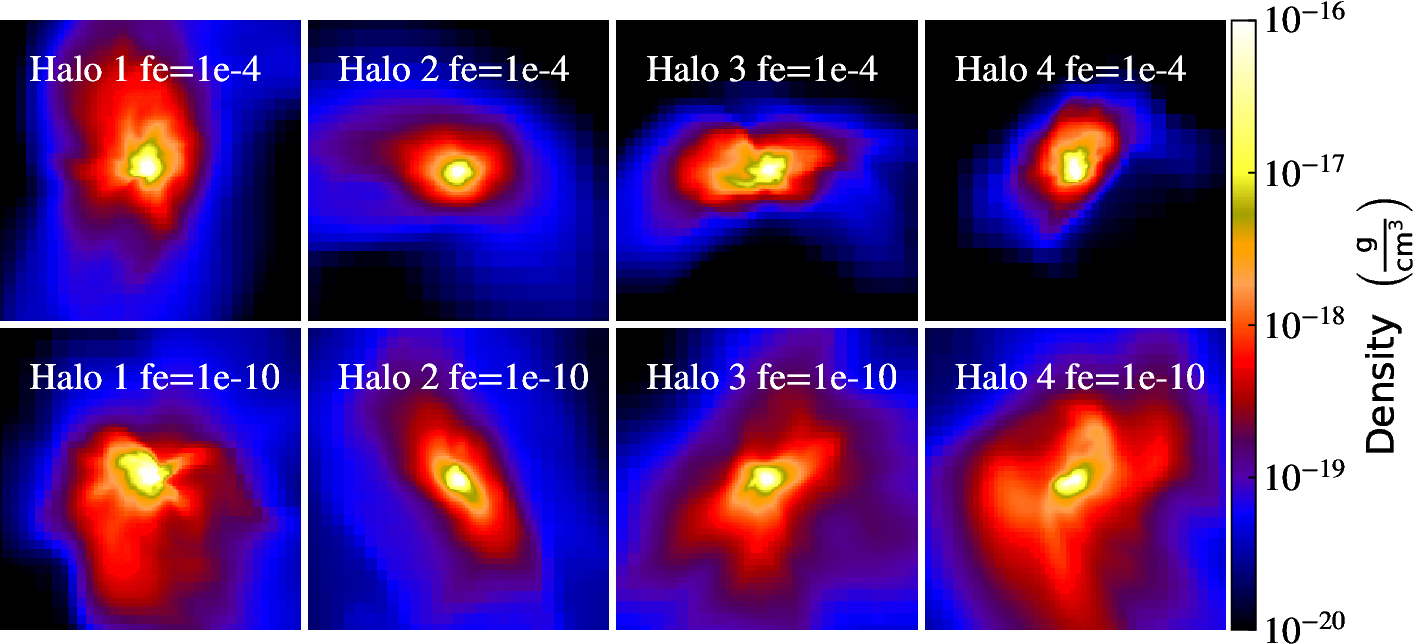,width=0.8\linewidth,clip=} \\ 
\epsfig{file=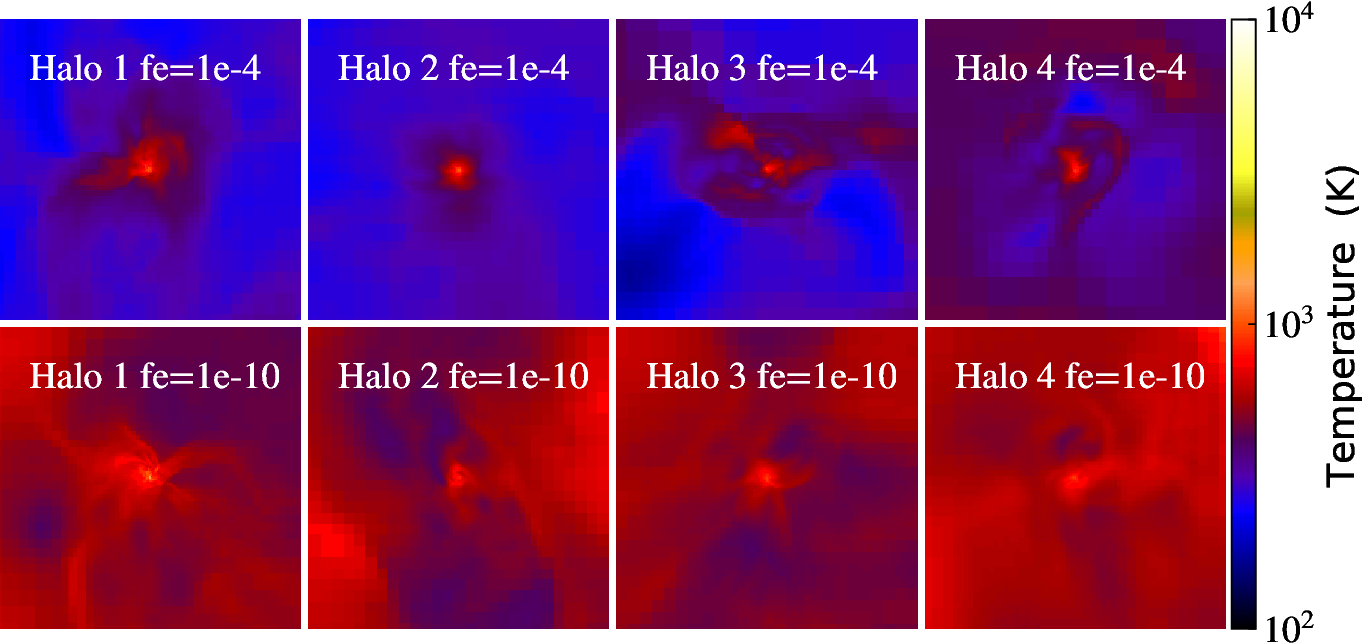,width=0.8\linewidth,clip=}  
\end{tabular}
\end{center}
\caption{The averaged gas densities and the mean gas temperatures along the line of sight for the the central 1 pc region are shown here for two $f_e$ cases($\rm 10^{-4}$ and $\rm 10^{-10}$).}
\label{fig4}
\end{figure*}

The average densities and temperatures in the center of the halos are shown in Fig. \ref{fig4} for the standard and $f_e=10^{-10}$ runs. The average densities in the central core are a few times $\rm 10^{-17}~g/cm^3$ and the mean temperatures of up to a 1000 K. In general, density structures are compact for the standard $f_e$ cases while more extended for $f_e=10^{-10}$ cases. Overall, disk like structures are observed in all cases due to rotation.
The gas temperatures on average are higher in the $f_e=10^{-10}$ runs compared to the standard cases. Such higher temperatures and densities lead to higher gas inflow rates mentioned above and are expected to foster the rapid formation of SMSs.

\section{Discussion and conclusions} 

We have conducted 28 cosmological simulations for four distinct halos to study the impact of different electron freeze out conditions. We here systematically varied the initial residual free electron fraction assuming  $f_e$ of $\rm 10^{-5}$, $\rm 10^{-6}$, $\rm 10^{-7}$, $\rm 10^{-8}$, $\rm 10^{-9}$, $\rm 10^{-10}$ and compared our results with the standard $f_e$ case.
Our findings show that halo collapse gets delayed for lower $f_e$ cases due to the lack of $\rm H_2$ formation and cooling, up to a few hundred Myrs for the lowest $f_e=$ cases ($\rm f_e \leq 10^{-8}$). The halos masses increase with initial lower $f_e$ runs and reach the atomic cooling limit for $\rm f_e \leq 10^{-8}$. 
Overall, the averaged densities and temperatures are higher in halos with lower $f_e$ runs compared to the standard case. 
Consequently, the enclosed gas mass in the halo centers increases from about $\rm 10^3~\Ms$ to $\rm 10^4~\Ms$ and the mass inflow rates from a few times $\rm 0.001~\Ms/yr$ to $\rm \sim  0.1~\Ms/yr$. For $\rm f_e \leq 10^{-8}$ cases, both the enclosed gas mass and the mass inflow rates are increased by an order of magnitude. Particularly, the mass inflow rates for the initial $\rm f_e \leq 10^{-8}$ cases are higher than the critical rate (0.04 $\rm \Ms/yr$) required for the formation of SMSs \citep{Hos13,Schleicher13,hle17,Woods19}.
Based on the large mass inflow rates and enhanced halo masses, we expect that such halos (with initial $\rm f_e \leq 10^{-8}$) may form $\sim 10^{4}$ M$_{\odot}$ BH seeds. 
Previous studies have shown that for such high inflow rates the feedback remains ineffective and SMSs are expected to collapse into similar mass BHs so-called direct collapse BHs \citep{Hos13,Schleicher13,Woods19,Nan23}. 

Our simulations have a resolution of about 40 AU and we find that disk like structures are formed.  While we find that disks are stable and may lead to the formation of a massive central object, fragmentation on scales smaller than our resolution cannot be ruled out. 
In fact, analytical disk models suggest that even if fragmentation occurs on smaller scales the clump migration timescale is shorter than the Kelvin-Helmholtz contraction timescale \citep{Inayoshi2014,Latif15v}. Therefore, clumps are expected to migrate inward and merge with the central object \citep{Latif15ad} and in addition to that, viscous heating in rapidly rotating disks may dissociate $\rm H_2$ at high densities \citep{Sch16}. Our arguments about the formation of massive objects in the presence of large inflow rates are also supported by radiation hydrodynamical simulations exploring fragmentation in massive pristine atomic cooling halos where collapse in the center is mediated by $\rm H_2$ cooling \citep{Regan18b,Latif21,Pat23}. They found that massive objects can still form through turbulent accretion and merging of clumps. 
Even if vigorous fragmentation occurs, it may lead to the formation of a dense Pop III cluster where stellar dynamical processes may lead to the formation of a massive BH seed of a few thousand solar masses \citep{2016MNRAS.457.2423Y,Reinoso18a,Sch22}.

Our findings reveal a new pathway for the formation of massive BHs which may naturally arise due to the free electron freeze out conditions in the early Universe. 
The BHs resulting from this scenario may explain the abundant massive BHs recently unveiled by the James Webb Space Telescope which are challenging to explain via conventional light BH seed models without invoking super Eddington accretion. Another interesting consequence of our work is that the delay in Pop III star formation in smaller mass haloes will lead to an increase in Pop II star formation at later epochs \citep{Johnson13}. This is mainly due to the first supernovae going off in bigger potential wells for lower $f_e$.

The prime focus of this study was to investigate the impact of different freeze out conditions on the formation of gaseous halos and study their physical properties.
In future work we will investigate the cosmological scenarios that can give rise to varying freeze out electron fractions and investigate their impact on the number density and demographics of SMBHs.

\section{acknowledgments}

MAL thanks the UAEU for funding via UPAR grants No.  12S111. SK acknowledges funding via STFC Small Grant ST/Y001133/1. For the purpose of open access, the authors have applied a Creative Commons Attribution (CC BY) license to any Author Accepted Manuscript version arising from this submission.

\section{Data Availability Statement}

The data in this study will be made available upon reasonable request to the corresponding author.

\bibliographystyle{mnras}
\bibliography{smbhs.bib}

\end{document}